\documentclass[12pt]{article}

\usepackage{a4wide}
\usepackage{amssymb}
\usepackage{bm}
\usepackage{latexsym}
\usepackage{amsfonts,amssymb}
\usepackage{graphicx,epsfig}
\usepackage{psfrag}
\usepackage{amsmath,amssymb}
\usepackage{mathrsfs}
\usepackage[T1]{fontenc}

\newcommand{\be}{\begin{equation}}
\newcommand{\ee}{\end{equation}}
\newcommand{\ba}{\begin{eqnarray}}
\newcommand{\ea}{\end{eqnarray}}
\newcommand{\ban}{\begin{eqnarray*}}
\newcommand{\ean}{\end{eqnarray*}}

\begin{document}
\begin{titlepage}
\pagestyle{empty}
\baselineskip=21pt
\vspace{2cm}
\begin{center}
{\bf {\Large Black-hole event horizons --- Teleology and Predictivity}}
\end{center}
\begin{center}
\vskip 0.2in
{\bf Swastik Bhattacharya${}^{a}$} and {\bf S. Shankaranarayanan${}^{b}$}
\vskip 0.1in
{\it ${}^{a}$ School of Physics, Indian Institute of Science Education and Research 
Thiruvananthapuram (IISER-TVM), \\ 
Trivandrum 695016, India} \\
{\it ${}^{b}$ Department of Physics, Indian Institute of Technology Bombay, \\
Mumbai 400076, India}\\
{\tt Email: swastik@iisertvm.ac.in, shanki@iitb.ac.in}\\
\end{center}

\vspace*{0.5cm}

\begin{abstract}
General Relativity predicts the existence of black-holes. Access to the 
complete space-time manifold is required to describe the black-hole. 
This feature necessitates that black-hole dynamics is specified by  
\emph{future or teleological} boundary condition. Here we demonstrate 
that the statistical mechanical description of black-holes, the {\it raison 
d'\^{e}tre} behind the existence of black-hole thermodynamics, requires \emph{teleological} boundary condition. 
Within the fluid-gravity paradigm --- Einstein's equations when projected on space-time horizons resemble Navier-Stokes equation of a fluid --- we show that 
the specific heat and the coefficient of bulk viscosity of the horizon-fluid 
are negative \emph{only} if the teleological boundary 
condition is taken into account. We argue that in a quantum theory of gravity, the 
future boundary condition plays a crucial role. We briefly discuss the possible implications of this at late stages of black-hole evaporation.
\end{abstract}


\end{titlepage}

\baselineskip=18pt

Physical laws, since the time of Newton, are predictive. Given an
initial condition, the physical laws predict the state of a system at a
later time. The two biggest revolutions in Physics of the
twentieth century --- Relativity and Quantum Theory --- are
predictive. The solutions to the dynamical equations subject to a
future boundary condition are viewed as the future influencing the
present and is generally deemed unphysical.

General relativity, while still predictive, brought in some
far-reaching changes. First major change, as compared to special relativity,  
is that the space-time is dynamic. Einstein equations can be solved subject to initial
conditions, however, the predictions are physical only after solving the 
space-time metric~\cite{ADM,MTW,Hawking-Ellis,Waldbk}. 
The second major change took place via the
concept of event horizon~\cite{MTW,Hawking-Ellis,Waldbk,Frolov}. 
To define the event horizon, one requires the complete knowledge of the
space-time manifold~\cite{MTW,Hawking-Ellis,Frolov}. Thus, the dynamics of the 
event horizon cannot be predictive and the existence of the horizon requires 
imposing future boundary condition on the horizon~\cite{Waldbk,Frolov,Membrane}.

The discovery of laws of black-hole thermodynamics, and the Hawking 
radiation~\cite{BCH-1973,Bekenstein,Hawking,Waldbk}, allowed
to identify, the surface gravity
of a black-hole with temperature and the area of the event horizon
with entropy. This also made clear that the
event-horizons via black-hole entropy play a crucial role in
identifying the microscopic degrees of freedom and hence, the
fundamental theory of Gravity~\cite{Parentani,Das:2007sj}.

Given that we do not yet have the complete description of quantum
gravity, there are two possible ways to go about obtaining the
microscopic description of the event horizon: First approach is to
assume the microscopic structures (like in Loop gravity or String
theory) and arrive at the entropy. Second approach --- 
\emph{pursued in this essay} --- is to map the gravity equations 
near the horizon to a familiar system for which the microscopic 
degrees of freedom are known.

It has been shown three decades ago that the Einstein's
equation projected on to the event horizon is similar to Navier-Stokes
equation~\cite{Damour,Price:1986yy,Padmanabhan:2010rp,Kolekar:2011gw,Membrane}. 
Thus, if energy and entropy of the fluid --- that are identical to that of the black-hole --- can
be obtained from the microscopic description of a fluid, it may be
possible to associate microscopic degrees of the fluid to the horizon. The
fluid contains two sets of parameters --- susceptibilities and transport coefficients. 
While the first set of parameters correspond to changes in local variables; the other set
involves fluxes of thermodynamic
quantities~\cite{Kadanoff,Kubo,Zwanzig}.  These parameters can be determined 
via statistical mechanical description of the fluid fluctuations. In
other words, treating the fluid not far from equilibrium, one can
study the statistical mechanical fluctuations from the
equilibrium and relate susceptibility/transport coefficient to the
autocorrelation function of a suitable dynamical
variable~\cite{Kadanoff,Kubo,Zwanzig}. 

Green-Kubo relations connect non-equilibrium processes
to the thermal fluctuations in equilibrium via fluctuation-
dissipation theorem~\cite{Kadanoff,Kubo,Zwanzig}. Mathematically,  
the change in the expectation value of any operator is linearly related to 
the perturbing source, i. e.,
\begin{equation}
\delta \langle {\cal O} (t) \rangle = \int dt' \chi(t - t')
\phi(t') \label{transport}
\end{equation}
where $\chi$ is the response (or Green) function and $\phi(t)$ is the
source. Assuming that the operator is Hermitian and the source is real
leads to the fact that response $\chi(t - t')$ must also be real.  The
Fourier transform of $\chi(t-t')$ can be written as:
\begin{equation}
\chi(\omega) = {\rm Re} \chi(\omega) + i \, {\rm Im} \chi(\omega)
\equiv \chi'(\omega) + i \chi''(\omega)
\end{equation}
The imaginary part of $\chi(\omega)$ can be written as :
\begin{equation}
\chi''(\omega) = -\frac{i}{2} \int_{-\infty}^{\infty} \!\!\!\! 
dt \, e^{i \omega t} \left[\chi(t) - \chi(-t) \right]
\end{equation}
$\chi''(\omega)$ represents the
part of the response function that is \emph{not invariant} under time
reversal. Evaluation of $\chi''(\omega)$ provides three key information
about the system. First, it provides information about the dissipative
processes. Second, it tells us whether the processes are causal or 
anti-causal~\cite{Onsager,1996-Evans.Searles-PRE}. 
Lastly, from Eq. \eqref{transport}, transport coefficients of normal fluids can 
be obtained once we know $\chi(\omega)$ in the hydrodynamic limit 
($\omega\rightarrow 0$)~\cite{Kadanoff,Kubo,Zwanzig}:
  \begin{equation}
   C_{\rm Transport} \propto \langle {\cal O}(t,\mathbf{r}) {\cal
     O}(0,\mathbf{r})\rangle, \propto \chi''(\omega = 0)
     \label{TransportCff-Gen}
  \end{equation}
where, $\langle {\cal O}(t,\mathbf{r}) {\cal O}(0,\mathbf{r})\rangle$
is the auto-correlation function of ${\cal O}$. 
Evaluation of the auto-correlation function again requires the causal (anti-causal) 
response of the horizon properties of the fluid. 

But, what does this digression have to do with black-holes? As we 
show in this essay, a similar analysis is possible for the
horizon-fluid and the issue of imposing the future boundary condition
comes to the fore. It also leads to a natural statistical mechanical
explanation of some of the puzzling aspects of black-hole
thermodynamics. A synopsis of the research programme advocated in this 
essay is provided in Fig. 1.

\begin{figure}[!htb]
\includegraphics[width=0.85\textwidth]{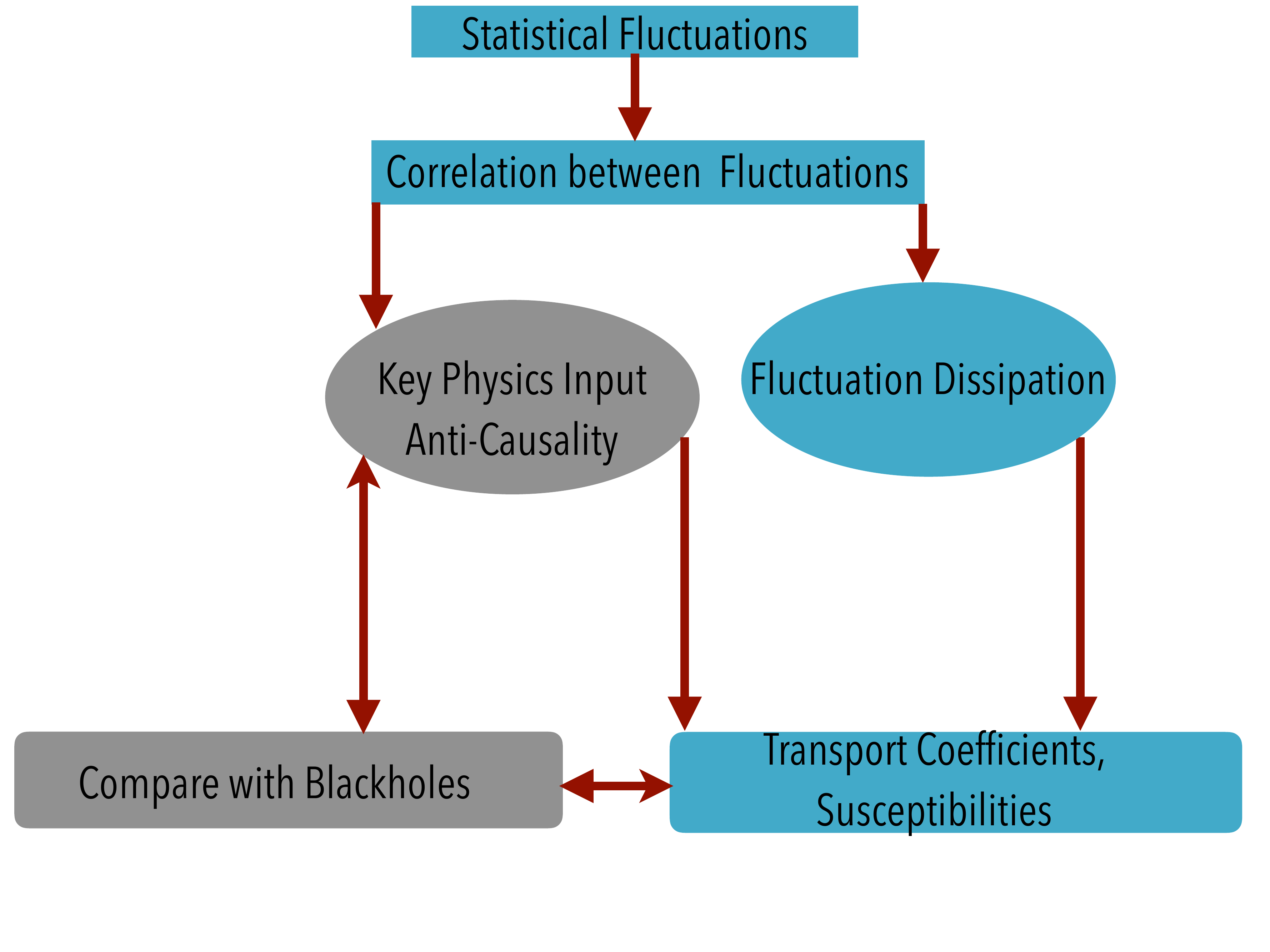}
 \caption{Bird's eye view of the programme and the importance of teleological boundary condition in understanding horizon dynamics}
 \end{figure}

As mentioned above, Einstein equations projected on to the event horizon of an
asymptotically flat black-hole leads to:
\begin{equation}
\label{eq:EinsProjected}
 \frac{D\theta}{dt}+\theta^2= -\frac{\kappa}{8\pi}\theta--\frac{1}{16\pi}\theta^2+\frac{1}{8\pi} \sigma_{AB}\sigma^{AB}+T_{\mu\nu}\xi^\mu\xi^\nu
\end{equation}
The above equation is similar to the energy conservation equation of a viscous fluid:
\begin{equation}
\label{eq:Fluid}
 \frac{\partial \mathcal{E}}{\partial t}+\mathbf{v.\nabla}\mathcal{E}= -\mathbf{\nabla .}[(\mathcal{E}+P)\mathbf{v}-\mathbf{v.\sigma'}]-\sigma'_{AB}\frac{\partial v_A}{\partial x_B} \, ,
\end{equation}
where $\mathbf{\sigma'}$ is the dissipative stress
tensor for the fluid given by, $\sigma'_{AB}= [v_{(A,B)}]_{Tracefree}+
\zeta \delta_{AB} \theta$, $\theta= \mathbf{\nabla.v}$, and
$\mathbf{v}$ is the velocity of the fluid. $\zeta (\eta)$ is the
coefficient of the bulk (shear) viscosity of the fluid.

Identifying $T = \frac{\kappa}{8\pi}$, $\zeta= -\frac{1}{16\pi}$,
$\eta= \frac{1}{16\pi}$, energy density of the fluid,
$\mathcal{E}=\theta$ and the momentum density of the fluid, $\Pi_A=
-\frac{1}{8\pi}n_a\xi^a_{;A}$ and $P=\frac{T}{4}$; equation 
(\ref{eq:EinsProjected}) is similar to the energy conservation equation 
of the fluid (\ref{eq:Fluid}) i. e.
       \begin{equation}
         \frac{D\mathcal{E}}{dt}+\theta \mathcal{E}= -\frac{\kappa}{8\pi}\theta+\zeta \theta^2+2\eta \sigma_{AB}\sigma^{AB}+T_{\mu\nu}\xi^\mu\xi^\nu
       \end{equation}
We note that 
the above equations are valid for an asymptotically flat space-time
with a single horizon. In this essay, our focus is to provide an understanding 
of the negativity of the bulk viscosity and show that 
the same can help to understand the negative specific heat of asymptotically flat 
black-hole space-times~\cite{Padmanabhan:1989gm,LyndenBell:1998fr} (See Fig. 1). 

To obtain statistical mechanical understanding of the transport coefficients of the 
horizon-fluid, we need to know how an external influence 
or source produces a response in the 
horizon-fluid~\cite{Skakala:2014eba,Bhattacharya:2014xma,Lopez:2015dlu,Bhattacharya:2015yga}. 
For simplicity, we focus on the horizon-fluid corresponding to
Schwarzschild black-hole~\cite{BV}, however, the analysis can be
extended for any asymptotically flat black-holes in General
Relativity~\cite{GBV}. The change in the area of the
horizon-fluid is given by the dynamics of the volume expansion
coefficient, $\theta^H$ of a congruence of null geodesics on the
horizon. It is governed by the Raychaudhury
equation, also the energy conservation equation for the
horizon-fluid~\cite{Bhattacharya:2015yga}. It can be cast in the form
of a linear response, by retaining only the volume change of the
fluid, given by,
\begin{equation}
 -\frac{d\theta^H}{dt}+ g_{_H} \theta^H \simeq  8\pi \mathcal{I}^H \label{response},
\end{equation}
where, $g_{_H} = 2\pi T$ and $\mathcal{I}^H$ is the source
which in this case is the energy flux. From the 
above equation, one sees that $\theta^H$ increases exponentially 
with time. Demanding that the horizon exists in the future necessitates 
to impose future boundary condition on $\theta_H$~\cite{Membrane}; imposing 
initial boundary condition on $\theta_H$ implies $\theta_H$ increases exponentially  
and destroys the black-hole. Since, $\theta_H$ is assumed to be small in deriving Eq. (8),
initial boundary condition can not be imposed. Thus the \emph{teleological boundary
  condition} can be viewed as a condition for the stability of the
black-hole event horizon or the condition that small external influences 
do not drive the fluid far from equilibrium~\cite{Bhattacharya:2015yga}. 

From Eq. (\ref{TransportCff-Gen}), the transport coefficients for the 
horizon-fluid can be rewritten as:  
  \begin{equation}
   C_{\rm Transport}= f(T,A) \langle {\cal O}(t,\mathbf{r}) {\cal
     O}(0,\mathbf{r})\rangle, \label{TransportCff}
  \end{equation}
where, $f(T,A)$, is a well-behaved function of $A$ and $T$, and
${\cal O}(t)\equiv \delta A(t)$. $\delta A$ is the change in the black-hole area and is proportional to $\theta_H$.  
From general properties 
of any fluid (including horizon-fluid) and using (\ref{response}), 
it can be argued that 
$\lim_{\epsilon\rightarrow0}\delta A(t)\sim \delta A_0(t)\exp[i(\omega-i\epsilon)t]$, 
i. e.  fluctuations in the area undergo damped oscillations in the hydrodynamic limit. 

For normal fluids in the presence of an external influence, the 
response is causal. However, as mentioned above, the response 
of an horizon-fluid to any external influence is anti-causal~\cite{Membrane}.
Mathematically, this implies: $\delta
A_{\rm Causal}(t) = \delta A(t)\Theta(t)$ and 
$\delta A_{\rm anti-causal}(t) = \delta A(t)\Theta(-t)$.  Hence, $C_{\rm Transport}^{\rm Anti-causal} = - C_{\rm Transport}^{\rm Causal}$~\cite{1996-Evans.Searles-PRE}. Thus, it can be shown 
that the bulk viscosity ($\zeta$) of the horizon-fluid is negative~\cite{BV,GBV}. 
A similar argument shows the specific heat of the horizon-fluid is negative~\cite{Bhattacharya:2017mrg}. 
It is important to stress that apart from the conventional techniques from the theory of fluctuations~\cite{Kadanoff,Kubo,Zwanzig}, the key physics input 
is the \emph{future or teleological} boundary condition in order to 
evaluate the statistical correlation of the change 
in the horizon area ($\delta A$) (See Fig. 1).

The analysis presented until now is semi-classical. We provide arguments that the teleological boundary condition plays a key role in the quantum theory. Assuming that the underlying theory for these fluctuations is a quantum theory, 
it is possible  to define an operator corresponding to this variable \cite{Bekenstein1,Areaop}. As long as the theory is in the space-time 
continuum,  our analysis clearly demonstrates the need to impose \emph{teleological boundary condition}. This is in stark contrast to all other physical theories.

While our analysis is for black-hole event horizons, it is possible to 
extend the analysis to  isolated horizons~\cite{Ashtekar}. For such horizons, the 
response is causal in nature, hence the coefficient of 
bulk viscosity is positive. However,  for the second law to be satisfied~\cite{Hawking-Ellis,Waldbk}, we need 
to impose future boundary condition. This condition makes sure that the black-hole 
horizon is stable at late times and is the famous Cosmic
Censorship Conjecture~\cite{Waldbk}.  
This seems to suggest that the future boundary condition is
integral to black-hole physics. One crucial question that remains is whether 
future boundary condition is  required at the late stages of black-hole evaporation. 
Our ignorance about the physics in this phase prevents us from saying anything 
concrete~\cite{Mathur,BMS}.  However, if the generalized second law remains valid, the horizon evolution might be governed by future boundary condition.

Ultimately the fundamental theory of Quantum Gravity might turn out to be very 
different from a theory based on a spacetime continuum~\cite{Raamsdonk}. It is interesting whether the future 
boundary condition plays a fundamental role there.

\noindent {\bf Acknowledgments} The authors thank S. Santhosh Kumar for the help with the flow-chart. The 
work is supported by Max Planck-India Partner Group on Gravity and Cosmology.

\end{document}